# *Rolling Polygons with Granular Material down an Incline*

Sidney Eckert, Phillip Ingalls, and J. West*

Indiana State University, Department of Chemistry and Physics

*To whom correspondence should be sent.

## Abstract

The present paper reports a novel behavior involving regular polygons with n sides and filled to varying degrees with granular materials. These are comprised of a set of hollow polygons produced on a 3D printer, and a single larger hollow hexagon fabricated with wooden sides and clear Plexiglas faces. Empty or full polygons stop immediately at shallow ramp angles, and roll the full length of the ramp for steep ramp angles approaching a terminal velocity. This is consistent with results previously reported by other investigators of rolling solid polygons and partially filled cylinders. In contrast, partially filled polygons released at shallow ramp angles accelerate to a "terminal velocity," but then come to an abrupt stop. The distance traveled is reproducible and dependent on the ramp angle, number of sides and the volume fill ratio and is minimized when filled to 0.4 of its total internal volume. For larger ramp angles, the partially filled polygons again approach a "terminal velocity," but the terminal velocity is minimized, again near the fill ratio of 0.4. A simple model is introduced, but it is successful only in replicating the overall trend in velocity as a function of time for the large angle cases.

# I. Introduction

The rolling motion of partially filled cylinders on ramps has been studied experimentally by previous investigators with granular materials,[1] and with liquids.[2-5] It is known that even shaking a closed can of soda affects the speed/acceleration of the cylinder.[4] In the cases of the granular filled cylinder, it was noted that a terminal velocity is approached, and that once steady state rolling is established, the granular material maintains a constant surface angle. Of particular interest here, is the report that the energy dissipation in the system occurs when the cylinder is filled to roughly 0.7 of its total volume. Similarly other investigators reported experimental results and theoretical modeling of the behavior of solid polygons rolling without slipping down a ramp (rolling hexagonal pencils,[6] and "grooved cylinders" which are polygonal gears.[6,7] In both cases, the polygons are reported to have approached a terminal velocity. For both of these studies, the ramp angles were restricted to very low angles (both less than 4 degrees) in order to maintain rolling motion rather than "bouncing."

The new stopping behavior reported here is observed with regular polygons, but in this case they are hollow, and filled to varying degrees with granular materials. Hence, the current paper combines elements of the experiments of references,[1,6,7] and to a lesser extent.[2-5] These studies indicate a general interest in the effective "rolling friction" of various objects and geometries.[8,9] When empty or filled completely, the behavior of the polygons is consistent with that reported for the solid polygons. However, for intermediate values of the volume fill ratio r, the behavior falls into three distinctly different qualitative cases: 1) The polygon stops at the first face impact; 2) the polygon approaches a terminal velocity as it proceeds to the end of the ramp; 3) the polygon seems to approach the terminal velocity, but then stops suddenly, over the course of a single or at most a few face impacts. This last behavior is found to be possible only within a small range of ramp angles, and only for partially filled polygons. The origin of this unexpected behavior might be related to the general observation that granular materials show large effects in behavior resulting from relatively small changes in conditions,[10-17] even small changes in temperature.[18,19] Other phenomenon involving granular materials have shown dramatic dependence on grain shape in particular.[20-23] It is somewhat surprising that the current study did not find significant differences in the rolling motion between the three different fill materials investigated.

In addition to the set of polygons produced with the printer, data from a single larger hexagon is also presented. This larger hexagon has clear Plexiglas faces allowing one to directly observe and video record the motion of the granular material within the hexagon as it rolls down the ramp. When examining the video recordings, the authors were unable able to identify any characteristic of the granular motion that distinguishes the stopping from the non-stopping behavior.

For a given ramp inclination, the data show that the distance a polygon travels is consistent, reproducible. For all polygons it ss found that the distance traveled is minimized when the polygon is filled to $r = 0.4$ of its full volume. For ramp angles steep enough that the polygons do not stop, the magnitude of the terminal velocity is also found to be minimized at roughly $r = 0.4$.

The position and velocity as a function of time is monitored frame by frame with a video recording system. A simple model based on conservation of angular momentum at each face impact is presented. The model is tested by comparing the predicted and measured velocity as a function of time for the smaller printed hexagon. The model uses only measured parameters of the system and does not include any adjustable parameters. It includes the effects of the granular material in a basic manner, and is similar to that of the work by Mead, et. al.,[24] a variation of which was derived independently by one of the authors (JW) as part of another study.[25] Qualitatively, the model predicts a smooth approach to terminal velocity Vt, but it does not predict the stopping behavior for any set of parameters tested. In addition, the model predicts that Vt should decrease monotonically with increasing fill ratio, contrary to the data obtained. The model therefore is clearly missing key physical effects. In particular, the model does not attempt to include the granular material at the grain level, and it is likely that the polygons do ont maintain continuous contact with the ramp. An improved model incorporating such elements could be illuminating.

The remainder of the paper is organized as follows. In Section II, the details of the experimental procedure are provided. The results are presented in Section III, while an outline of the theoretical model and some of the predicted results are presented and compared to the experimental data in Section IV. Finally, the conclusions are presented in Section V.

## II. Experimental Details

The experiment utilizes a set of hollow box versions of regular n-sided polygons (n-gons) with n = 6, 8, 10, and 18. These are designed in a CAD program and printed on a hobby level 3D printer (MakerGear M2), using standard Polylactic Acid (PLA) printer filament (Hatchbox part # 3D PLA-1KG1.75-WHT). Each polygon has an inside center-to-vertex radius R = 3.36cm, height H = 6.2 cm, and volume V = 210 $cm^3$. Each box sidewall is 0.20 cm thick, or composed of two overlapping box-and-lid walls each 0.10 cm thick. Each end face is 0.10 cm thick. The mass of each polygon is M = 0.033 kg, composed of an off-white, opaque ABS plastic. A large black spot is drawn on the lid face of each polygon with a permanent black marker to facilitate video tracking.

A single larger hexagon solid wooden sidewalls 1.2 cm thick was also used. Each face is Plexiglas cut from a clear 1/8" sheet, is attached to the main body with simple wood screws. This hexagon has an inside diameter face to face = 0.113 m, outside edge length = 0.080 m, inside edge length 0.066 m, and height 0.096 m. The outside volume = $1.60 \times 10^{-3}$ m$^3$, and inside volume = $1.09 \times 10^{-3}$ m$^3$ (roughly 1100 cm$^3$). The total mass of the larger hexagon including Plexiglas and screws is $M_{Hex}$ = 0.2428 kg. A large black dot is drawn on one face to facilitate video tracking, and the clear faces allow the motion of the granular materials to be observed and video recorded as it rolls.

The polygons are filled to various volume fill ratios r (r = 0, 1/8, 2/8,...8/8) with three different granular materials: roasted salted green peas (roughly spherical, dimeter = 0.8 cm, somewhat irregular in size and shape), uncooked white rice, and Airsoft or Match Grade .20g Precision BBs (double polished, seamless, diameter = 0.600 cm, model 20GPW2J). Each polygon is filled successively with each material, and the volume of the material Vo is measured using a graduated cylinder. The cylinder is then filled to the volume desired V based on the fill ratio r, where V = r*Vo.

Two ramps are used in the experiments. The first is a solid wooden ramp (1.210 m x 0.292 m x 0.010 m). The second is an aluminum air track originally used in classroom kinematics demonstrations. The ramp is roughly 20 years old, and the model number is not known. The air track is in the shape of a box with protruding support beams, giving an I-beam cross-section shape as shown in Fig. 1. The air track is 2.539 m long, and the I-beam shape gives the ramp a channel 7.9 cm wide and 2.50 cm high. The small polygons fit into this channel and roll freely.

To prevent slipping, the wooden ramp surface is covered with a single layer of Uline anti-slip tape 2" x 60' (model S-7187), and the aluminum air track surface is covered with a single layer of Norton 400 Grit PSA Sandpaper (4" x 25yds, Dura Block Gold Sticky Back Roll).  The ramps are elevated at one end via a combination of up to 3 lab jacks ( Boekel LAB-JACK No. 19089-001) and up to 3 reams of paper (500 sheets each) or soft back textbooks for shock absorption.  The middle of each the ramp is also supported in the same manner to minimize ramp oscillations.  The angle of the ramp $\theta$, as measured from horizontal is determined via basic trigonometric relationships using the measured height and length of the ramp.

The polygon is supported by hand on a vertex at the top of the ramp and released from rest.  The polygon position vs time is recorded with a Logitech HD C615 webcam controlled with Logitech Webcam software run from a Windows based laptop computer.  The position and time data were obtained via a frame-by-frame logging of the position of the central black colored onto the center of each polygon face using Logger Pro 3.12 software.

## III. Results

For small ramp angles (in most cases less than 10 degrees), the polygons stop at the first face impact.  For larger ramp angles ($\theta > 17$ degrees) all of the polygons accelerate and approach a "terminal velocity" as they proceed to the end of the ramp.  The approach to terminal velocity is consistent, with the results for solid polygons found in previous studies studies.[1,6,7]

For intermediate values of r and $\theta$, the polygons exhibit an initial acceleration, seem to reach the terminal velocity, but then abruptly stop after rolling a reproducible distance D.  The authors are not aware of any previous results by other researchers indicating similar behavior. Over the range that this behavior is observed, D increases linearly, according to D = Do + J*$\theta$, where J is in units of meters/degree and J = 0.46 rice; 0.57 BBs; 0.63 peas (see Fig. 2).

For a given angle, it is also found that the distance traveled before stopping is minimized at roughly r = 0.4 for each polygon and each fill material. The experimental values of D as a function of r for the small hexagon at a ramp angle θ = 20.6 degrees are shown in Fig. 3. Likewise for a larger ramp angle, the value of Vt is also found to be minimized for r = 0.4, but notice that the curve is not symmetric about that point.

An attempt was made to fit the data with a model, similar to that found in references

For r in the range r = 0.3 – 0.7, there is a significant amount of open space inside the polygon allowing more active grain motion and more violent grain-grain and grain-boundary collisions. Much of the grain at the surface is actively involved in the recurring wave-like motion. The rest of the grains remain largely inactive at the bottom of the hexagon during one cycle of rotation, the remainder. The images of the rice grains from inside the large rolling polygon are strikingly similar to those seen in Figs 2 and 3 in the work by Avila, et. al.[17] and reported in additional work.[26,27]

The motion of the granular material during the rolling motion is very regular for all cases observed. Which is to say that no "distinctive" motion of the granular material is identified that allows one to predict which face impact will be the last face impact, right up until the point that the polygon stops. A close examination of the velocity data values seems to indicate a slight **increase** in the hexagon velocity occurs a few face impacts before it stops. Two examples are shown in Fig. 4. The arrows indicate where increase is observed.

## IV. Theory

Consider the motion of a regular polygon rolling down a ramp without slipping ramp (see Fig. 5) with vertices labeled A, B, C and so on. If the speed of the polygon is low enough, it will rotate about each successive vertex, with a face-to-surface collision at each rotation axis transitions. This assumption is not always valid. For "large" ramp angles or large polygon speeds, even the partially filled polygons will "walk" on the vertices, rather than impacting with the full face.[7,24] No such walking behavior was noted for the data presented in this paper.

In rotating about a single axis from one face impact to the next, the potential energy of the hexagon will decrease by an amount Mgh, where M is the mass of the polygon, g is the usual 9.81 m/s$^2$, and h is the change in the center of mass altitude (h = side length *sinθ). This leads to an increase in the rotational kinetic energy of the hexagon about the axis of rotation taken here to be the direction of the positive z-axis.

At the face impact, the rotation axis is shifted to the next successive vertex, but remains along the z-axis. This face impact is accompanied by a significant loss of mechanical energy. The model is based on the assumption that the angular momentum of the hexagon, as measured about the **second** vertex, is conserved during this impact event.[24,25] At that time it is rotating about vertex A with angular velocity $\Omega_A$. The total angular momentum of the polygon about vertex A can be broken into a "spin" component,[28] and an "orbital" or "translational" component, written as

$$\vec{L}_A = Io\vec{\Omega}_A + M\vec{R}_A \times (\vec{\Omega}_A \times \vec{R}_A) = \left[(Io + MR^2)\Omega_A\right]\hat{z}, \quad (1)$$

where Io is the moment of inertial about the axis of symmetry, and **R**$_A$ is the vector from vertex A to the center of mass. At the next face impact, the polygon will begin rotating about vertex B. The magnitude angular momentum L$_B$ as measured about vertex B is always less than L$_A$ because the translational motion of the center of mass about A will have a component that is not perpendicular to R$_B$.

The angular momentum of the polygon about vertex B will be conserved during the "face collision" at vertex B. The spin component of the angular momentum is the same as when measured about vertex A, but the translational component now depends on the distance to the center of mass as measured from vertex B. Conserving angular momentum about vertex B provides a direct relationship between the angular velocity of the polygon after the collision and before the collision

$$\vec{L}_B = Io\vec{\Omega}_A + M\vec{R}_B \times (\vec{\Omega}_A \times \vec{R}_B) = \left[(Io + MR^2)\Omega_B\right]\hat{z}, \qquad (2)$$

The value the angular velocity about vertex A immediately before impact is determined directly from conservation of mechanical energy, or from numerical integration of the rotational equation of motion. The value of M is directly measured, and Io is calculated analytically in terms of M and the measured dimensions of the polygon. The expression in Eq. (2) implies that a certain **fraction** of the kinetic energy will be lost at each face impact: $KE_B/KE_A = Q_n < 1.0$. The value of $Q_n$ is conceptually similar to that of the coefficient of restitution for bouncing objects,[28] and is equivalent to the parameter "$\varepsilon$" in reference 6. For n = 3 (equilateral triangles), Eq. (2) predicts that the angular velocity about vertex B is negative. That is another way of saying "regular triangles do not roll." For n > 3, we note that Q > 0 and that Q increases rapidly with increasing n. For example: $Q_4 = 0.06$, and $Q_6 = 0.42$.

This still indicates and energy loss at each face impact. For small θ, the remaining rotational KE does not allow the center of mass to rotate about vertex B beyond the point above that vertex, and the polygon simply rotates back to the AB face and ceases its motion. On the other hand, for steeper ramp angles, the kinetic energy remaining after the face impact will allow the polygon to rotate far enough about vertex B that the torque acting on the center of mass becomes positive, and the polygon successfully rolls to face BC. If that is possible, then the polygon necessarily has enough kinetic energy after every face impact to keep repeating this pattern to the end of the ramp, approaching a terminal velocity as shown in Fig 6. At terminal velocity Vt, one would expect that $(1-Q_n)*KE = Mgh$.

This simple model **does not** predict the possibility of a solid polygon stopping once it starts rolling down the ramp. Either it stops at the first face, or it necessarily continues to approach Vt. A slightly more complete model represents the polygon shell, and the granular fill material as separate component. Note that the video analysis of the grains in the large hexagon shows that at a face impact event, the grains undergoe a sudden re-distribution such that the surface of the grains is reset parallel with the horizon. In addition, between impacts the grains remain effectively immobile relative to the polygon. This differential behavior of grains in rotation objects is also well known.[29-31] Treating the grains as a solid object between impact events allows one to subdivide the grains into a set of distinct triangular segments of mass $\{m_k\}$. The use of the triangular cross section allows for the mass $m_k$, location of the center of mass $\mathbf{R}_k$, and the moment of inertia $I_k$, of each grain segment to be completely determined by the coordinates of its vertices. Those vertex coordinates are completely determined by the fill ratio r, and the ramp angle $\theta$ (see Appendix for details). The mass M, moment of inertia I, and center of mass $\mathbf{R}$, of the combined polygon shell and grains are defined as

$$I = I_{poly} + \sum_k I_k$$
$$M = M_{poly} + \sum_k m_k \qquad (3)$$
$$\vec{R}_{cm} = \left[ M_{poly} \vec{R}_{poly} + \sum_k m_k \vec{R}_k \right] / M$$

where "poly" indicates values from the hollow shell.

For a hexagon, up to r = 0.738, it is only necessary to use either two or three grain segments (a full hexagon needs 5). It can also be shown that any regular polygon can be subdivided in the same manner. Conceptually, the model is still based on the assumption that the total angular momentum is conserved about the new axis of rotation at each face impact. By using the center of mass coordinates of the shell and triangular grain segments to determine the net torque, the dynamics of the polygon between face impacts are modeled with the rotational version of Newton's Second Law. The model has no adjustable parameters. There is now an **additional** energy loss mechanism. At a face impact, the center of mass of the grains drops an appreciable distance, but there is not an associated gain in kinetic energy. That amount of potential energy is lost from the system in the form of grain-grain interactions.

Calculations of position vs time are performed using a program written in Visual Python (VPython-Win-32-Py2.7-6.11). The initial conditions of the grain surface are chosen to match the experiment. The model predicts that for the small hexagon for n = 6, r = 3/8 the hexagon will not rotate beyond the first impact for ramp angles of 12º or less, in fair agreement with the results presented here. However the model and experimental results obtained here do not agree with those reported for solid hexagons, which indicate rolling rather than stopping for angles as low as 4º by previous investigations.[6,24]

The model **does** indicate that the composite objects approach a terminal velocity for larger ramp angles (See Fig. 7) without any adjustable parameters, but does **not** reproduce the stopping behavior for any combination of n, r or $\theta$. As shown in Fig. 6, at a ramp angle of $\theta$ = 20.6º the model also fails to produce the minimum in Vt as r a function of r. Perhaps a model that includes a bouncing element to the motion, or a model that deals with the grains in a detailed manner would be more successful.

## IV. Conclusions

An investigation of the rolling motion of regular polygons partially filled with granular material, rolling freely down a ramp after being released from rest is presented. A new stopping phenomenon is reported, which occurs only for partially filled polygons, but only within a narrow range of ramp angle. For those conditions, the polygon accelerates as if approaching a terminal velocity, but then abruptly stops after traveling a reproducible distance D. The value of D found to have a linear dependence on θ within the allowed range, and D is minimized at for r = 0.4, with a slight dependence on the material used being evident. An analysis of the velocity in the stopped cases suggests that there may be a slight increase in velocity for a few face impacts immediately before the polygon stops. Using a visual frame-by-frame inspection of the video of rice grains within a larger hexagon, the authors are not able to discern a pattern in the wave-like motion of the granules within the polygon associated with this abrupt stop. It is also found that for larger ramp angles the polygons approach a terminal velocity Vt, and that Vt is minimized for r = 0.4. A simple model is presented that fits the qualitative velocity vs time behavior without any adjustable parameters, but clearly the model is not sophisticated enough to account for the observed stopping behavior. The authors are currently investigating whether a model which treats the grains in more detail, or which considers the possible bouncing of the polygons at impact might meet with more success. It might prove useful to measure the force required to roll the polygons across a flat surface, or to measure the force required to hold the polygons in place while on a conveyor belt mechanism.

## Acknowledgements

The authors gratefully acknowledge financial support of the Indiana State University Department of Chemistry and Physics, and the Center for Student Research and Creativity for financial support. The authors also thank Shelley Arvin (ISU Library), and Michelle Baltz-Knorr (ISU Physics Laboratory Coordinator), and Tyler Jenkins (Physics Major).

**Appendix: Triangle Properties from Vertex Values**

It is possible to determine the location of the center of mass, the area (and mass), and the moments of inertia of a uniform density triangular prims using only the coordinates of its vertices. The authors feel that while this information can be found, or at least compiled from published work, it could prove useful to have the information conveniently available for research and educational uses. By dividing the grain segments into triangles, it is possible to determine the mass, center of mass location, moment of inertia, angular momentum and kinetic energy of each segment individually.

Consider a triangle with vertices labeled A, B, and C at the coordinates { $(X_A, Y_A)$, $(X_B, Y_B)$, and $(X_C, Y_C)$ }. By basic geometric argument, the center of mass of a triangle must be at the intersection of the bisectors of that triangle. One can easily show using the line-intercept formula that the coordinates of that intersection point must be the average value of each of the coordinates

$$x^* = \frac{X_A + X_B + X_C}{3} \quad \text{and} \quad y^* = \frac{Y_A + Y_B + Y_C}{3}. \tag{A3}$$

It is also well known that the area of any triangle equals one half of the area of a parallelogram constructed using two of its sides. The directed area **T**, of a parallelogram is in turn defined, independent of the coordinate system chosen, by the cross-product of the vectors associated with any two adjacent sides of the parallelogram, so that

$$\vec{T} \equiv (\vec{B} \times \vec{C}), \tag{A4}$$

where **B** and **C** are defined as the vector from A to B and A to C, respectively,

$$\vec{B} = (X_B - X_A)\hat{e}_x + (Y_B - Y_A)\hat{e}_y$$
$$\vec{C} = (X_C - X_A)\hat{e}_x + (Y_C - Y_A)\hat{e}_y \tag{A5}$$

Combining Eqs. A4 and A5, gives an expression for the area of any triangle in terms of its vertex coordinates

$$\text{Area} = |(X_B - X_A)(Y_C - Y_A) - (Y_B - Y_A)(X_C - X_A)|/2. \tag{A6}$$

If the triangle vertices are chosen in a counter-clockwise sense, the area given by Eq. A6 is always positive.

The expression in Eq. (A6) is very helpful as the area is proportional to the mass of the triangle. For a given choice of r and ramp angle, the expression in Eq. A6 is used to determine the number of triangle segments needed for the fill material, as well as the vertex locations of those triangles for the fill ratio of interest by requiring the sum of the areas of the grains be equal to the fill ratio times the Polygon Total Area.

In deriving expressions for the moment of inertia, begin by choosing the axis in the plane of the triangle that makes the integration of the mass elements the easiest to visualize. Such a choice is shown in Fig. A1. By direct integration, one can find Ix and Iy, and then by use of the Perpendicular Axis theorem and the Parallel Axis theorem, one can find the moment of inertia about the center of mass

$$I_{CM} = \frac{HD^3 - HKD^2 + HK^2D + DH^3}{36} = \text{Area} * \frac{D^2 + K^2 + H^2 - KD}{18}. \tag{A7}$$

So if $I_{CM}$ is known, and the location of the center of mass is known, then finding the moment of inertia about any other point is simply a matter of using the parallel axis theorem a second time. Expressions for D, K, and H for arbitrary orientations of the triangle in terms of the vertex coordinates are still needed and can all be found from the basic triangle geometry as

$$D \equiv |\vec{B}|, \quad K \equiv \frac{\vec{B} \bullet \vec{C}}{|\vec{B}|} \quad \text{and} \quad H \equiv (C^2 - K^2)^{1/2} = \frac{|\vec{B} \otimes \vec{C}|}{|\vec{B}|}. \tag{A8}$$

By using the definitions of the dot and cross products the relations in Eq. (A8) can be written explicitly in terms of the coordinates of the triangle as

$$D = \left[(X_B - X_A)^2 + (Y_B - Y_A)^2\right]^{1/2}$$
$$K = \frac{(X_B - X_A)(X_C - X_A) + (Y_B - Y_A)(Y_C - Y_A)}{\left[(X_B - X_A)^2 + (Y_B - Y_A)^2\right]^{1/2}}$$
$$H = \frac{(X_B - X_A)(Y_C - Y_A) - (X_C - X_A)(Y_B - Y_A)}{\left[(X_B - X_A)^2 + (Y_B - Y_A)^2\right]^{1/2}}$$

(A9)

These results then allow one to calculate the location of the center of mass (Eq. 3A), the area (Eq. A6), and the moment of inertia about the center of mass (Eqs. 7A – 9A).

Fig. 1. The air track is 2.539 m long, and the I-beam shape gives the ramp a channel 7.9 cm wide and 2.50 cm high. The small polygons fit into this channel and roll freely.

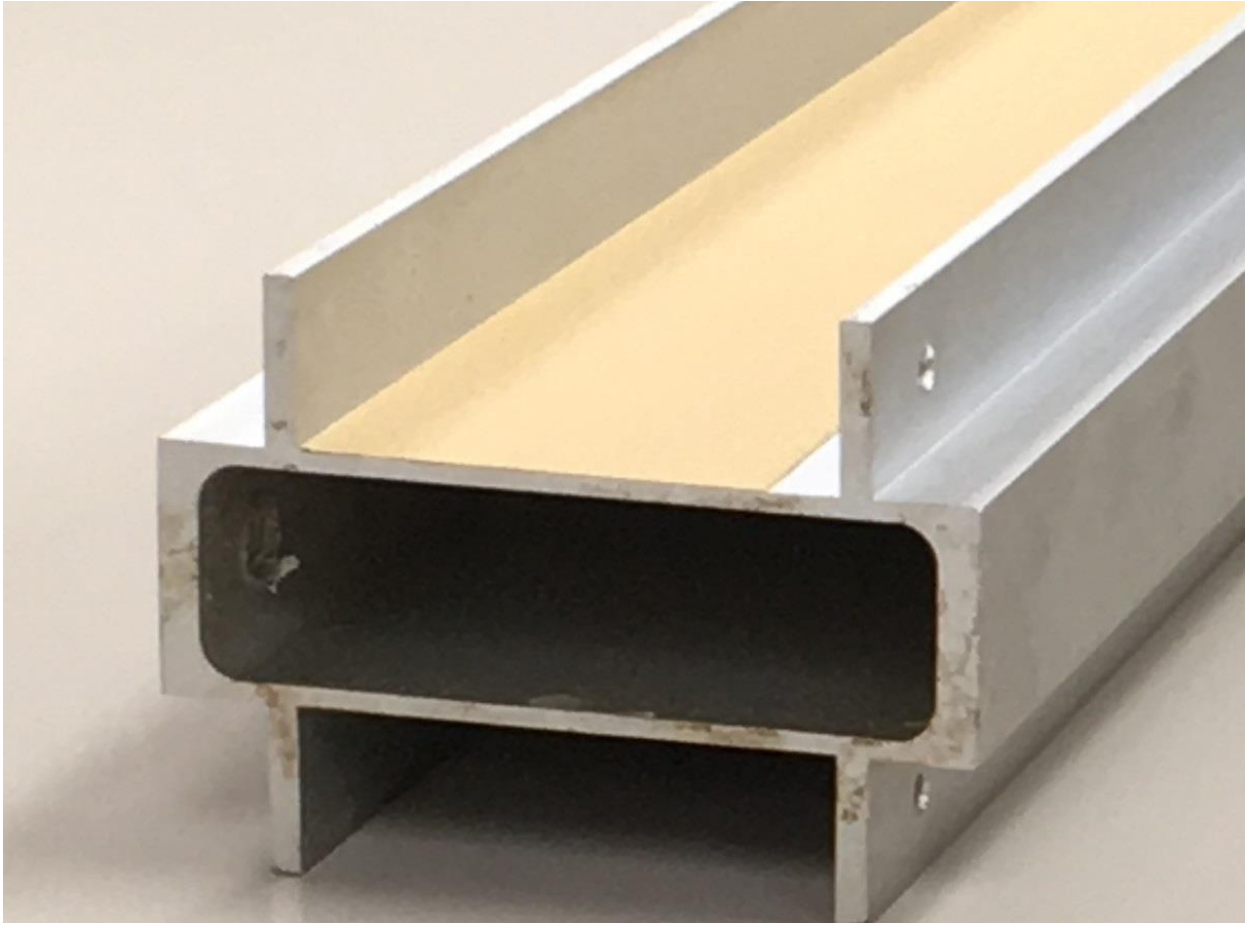

Figure 2. There is a strong linear correlation between the angle and the distance traveled for a partially filled polygon. Here the data is for an octagon-box (n = 8), fill ratio of r = 0.375, and with three different fill materials. Distances are averages, with an uncertainty of plus or minus one box side, in this case +/- 0.02 m.

A) Airsoft BBs, fit D = (0.60 m/deg) θ – 6.61 m, $R^2$ = 0.95.

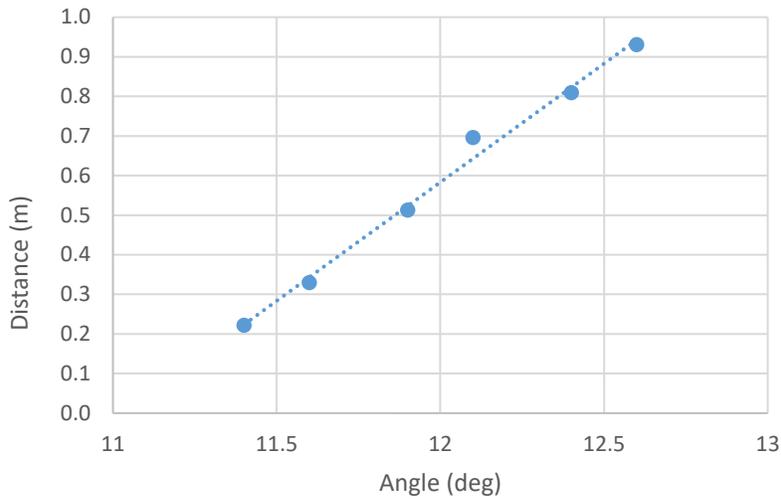

B) Rice, fit: D = (0.46 m/deg) θ – 5.62 m, $R^2$ = 0.95.

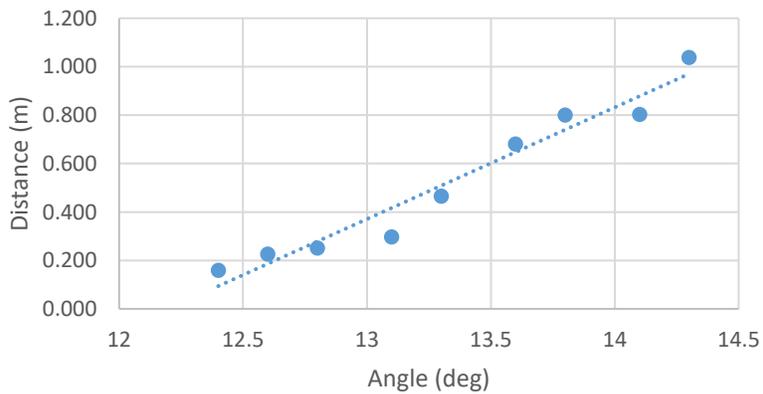

(Fig. 2C Next page)

C) Green salted peas, fit: D = (0.63 m/deg) θ – 5.20 m, $R^2$ = 0.91.

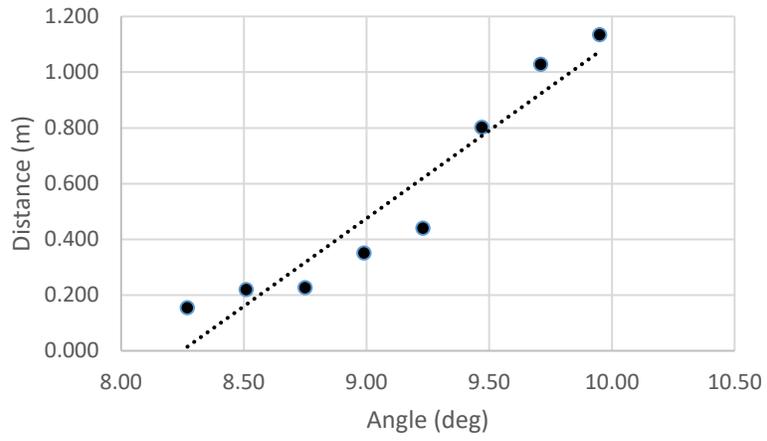

Fig. 3. The experimental data shows that the terminal velocity of the small hexagon (n = 6), on a ramp at angle 20.6 degrees has a distinct minimum near r = 0.40.

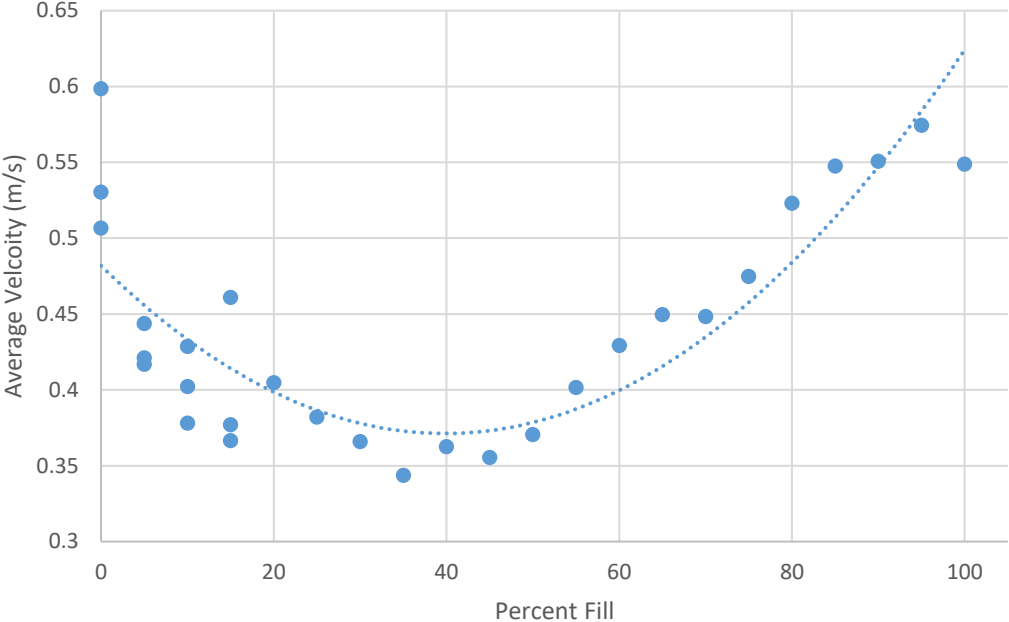

Figure 4. Under conditions for which the polygons stopped, the authors noted a possible recurring phenomenon of an increase in the velocity of the polygon in immediately preceding the abrupt stop. Two examples are shown here: A) n = 6, r = 0.357, theta = 18.4 degrees; and B) n = 6, r = 0.500, theta = 14.5 degrees

A)

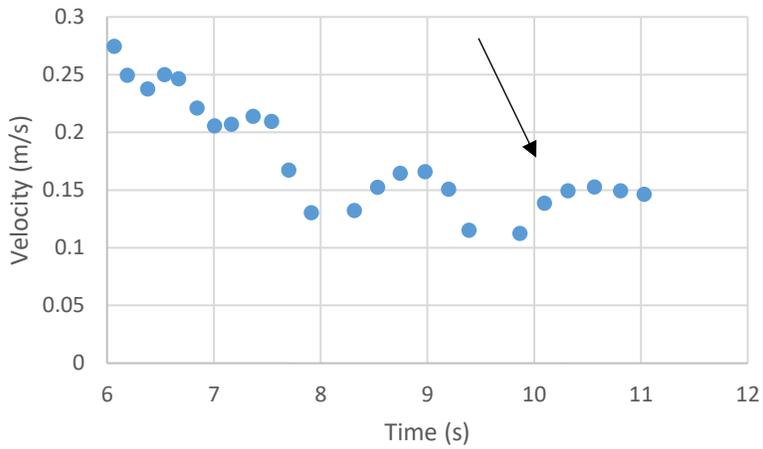

B)

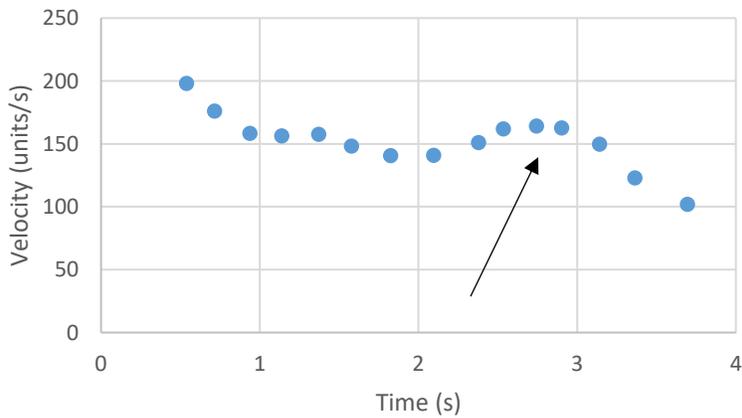

Fig. 5. The grains inside the polygon at each face impact become parallel with the horizon. For the rotation about the next vertex, the grains remain in place relative to the polygon until the next face impact. At that point, the impact causes largely liquid-like motion, allowing the granular material to "slosh" back to a surface roughly parallel to the horizon. It is presumed that the potential energy of the "Wave Mass" (mw) is lost to the mechanical system via grain-grain friction interactions.

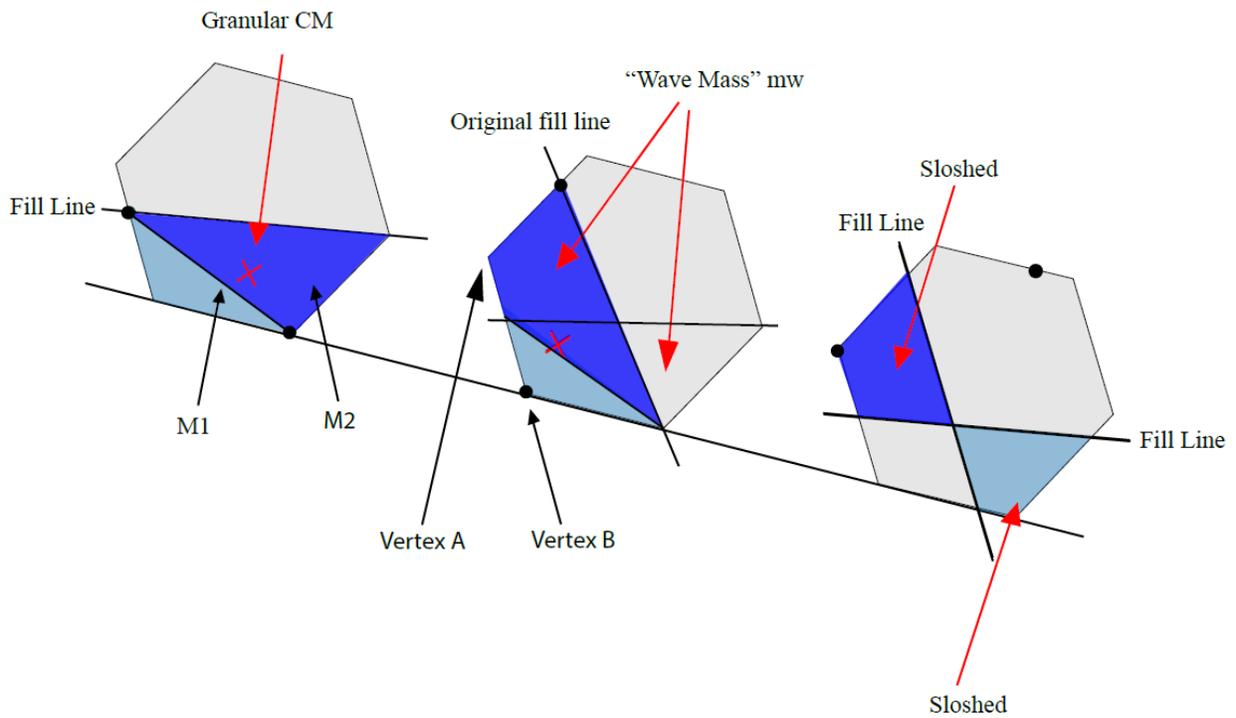

Fig. 6. Model prediction of terminal velocity Vt as a function of fill ratio r. The trend shown does not agree with the experimental results shown in Fig. 3.

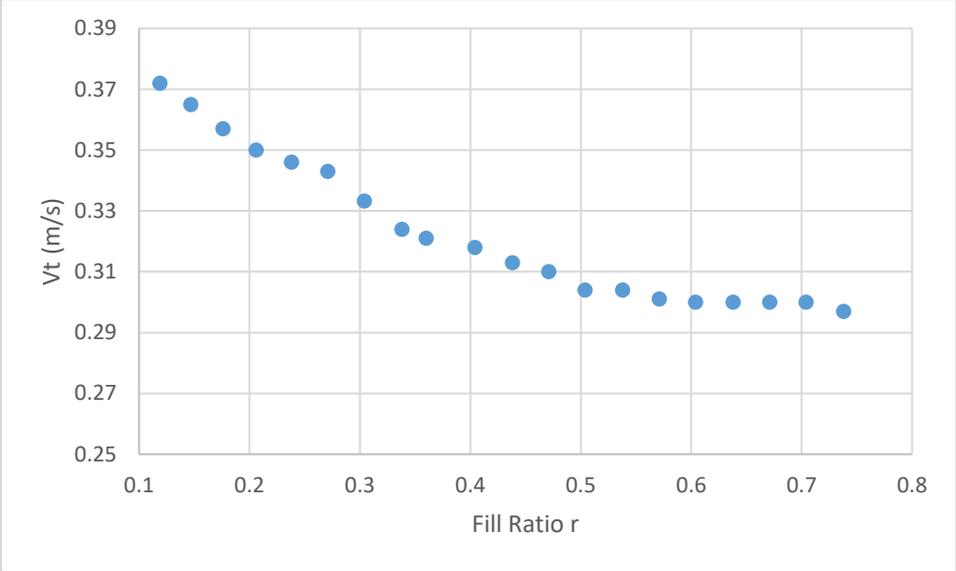

Fig. 7. The velocity as a function of time as predicted from the computer program written to model the dynamics of the partially filled polygons. The values shown are for parameters consistent with the small 3D printed hexagon-box, filled to r = 0.25, released from rest on an incline of 20 degrees. Notice the smooth approach to a terminal velocity. Top is model, bottom is data.

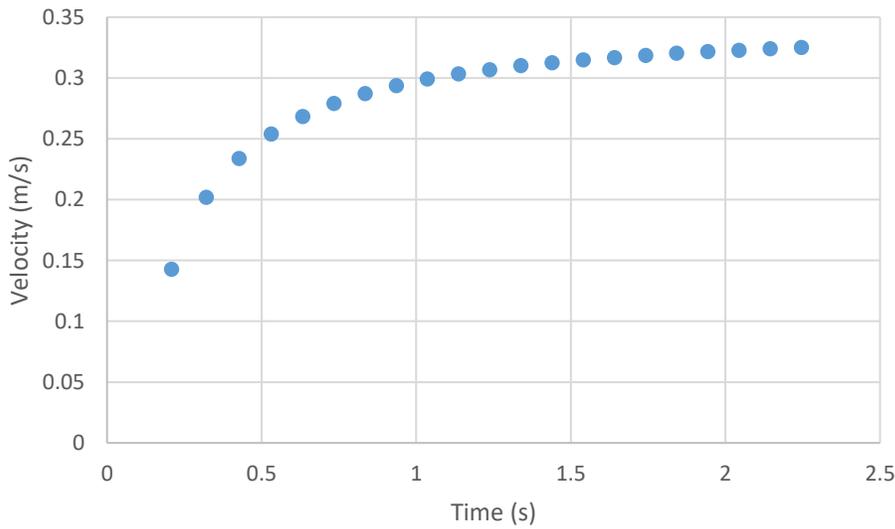

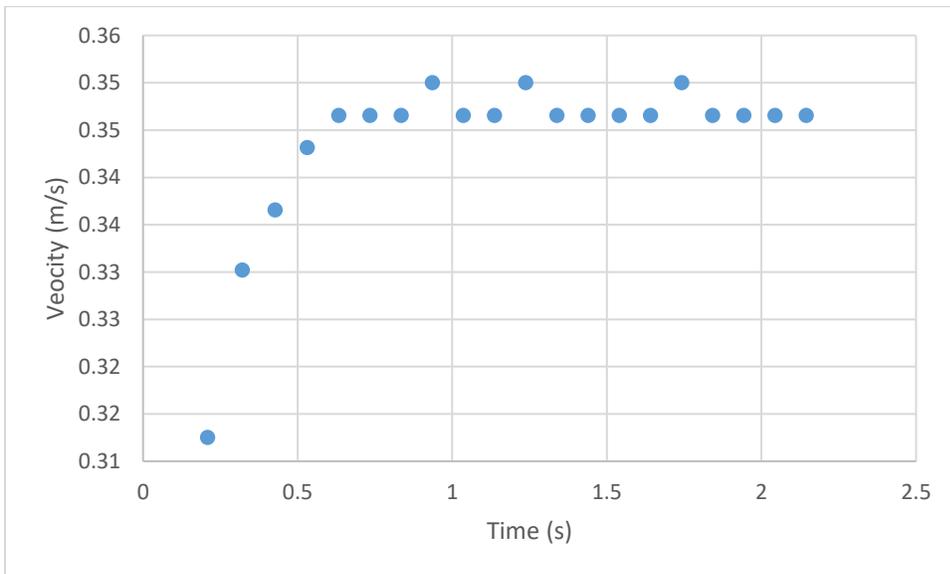

Fig. A1. On choice of vertex coordinates chosen to illustrate the definitions used in calculating the area, the moment of inertia about the center of mass, and the coordinates of the center of mass of a uniform triangular prism.

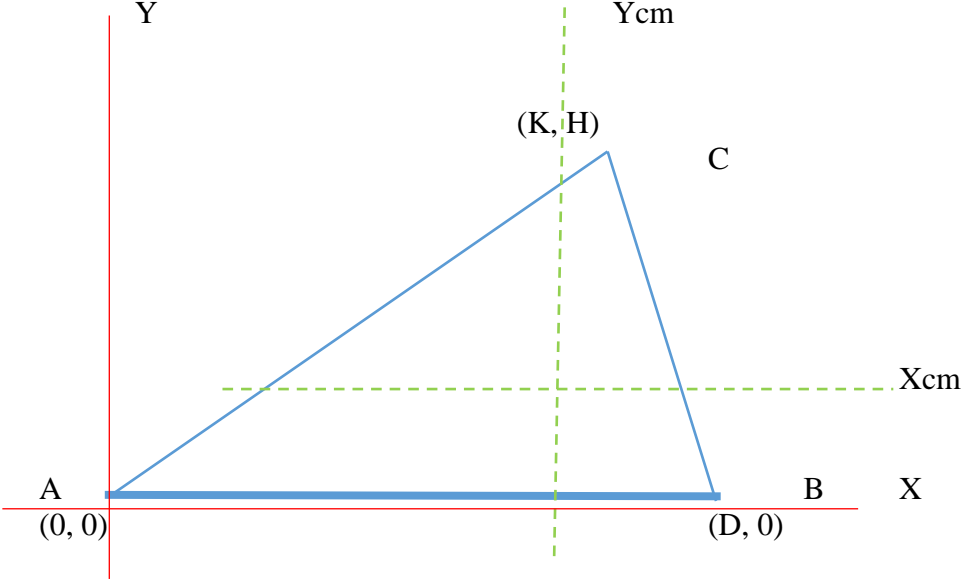